\documentclass[12pt]{iopart}

\usepackage{amssymb}
\usepackage{bm}
\usepackage{graphicx}

\newcommand{\cgc}[2]{ \left<#1 | #2 \right>}

\begin{document}

%
%
\title[Atomic ionization by twisted photons: Angular distribution of emitted electrons]{Atomic ionization by twisted photons: \\ Angular distribution of emitted electrons}

%
%

\author{O~Matula$^{1,2}$, A~G~Hayrapetyan$^{1}$, V~G~Serbo$^{3}$,\\ A~Surzhykov$^{4}$ and S~Fritzsche$^{4,5}$}

\address{$^1$ Physikalisches Institut, Universit\"at Heidelberg, D--69120 Heidelberg, Germany}
\address{$^2$ GSI Helmholtzzentrum f\"ur Schwerionenforschung, D--64291 Darmstadt, Germany}
\address{$^3$ Novosibirsk State University, RUS--630090 Novosibirsk, Russia}
\address{$^4$ Helmholtz--Institut Jena, D--07743 Jena, Germany}
\address{$^5$ Theoretisch--Physikalisches Institut, Friedrich--Schiller--Universit\"at Jena, D–-07743 Jena, Germany}

\ead{omatula@physi.uni-heidelberg.de}

%
%
%
%

\begin{abstract}
We investigate the angular distribution of electrons that are emitted in the ionization of hydrogen--like ions by twisted photons. Analysis is performed based on the first--order perturbation theory and the non--relativistic Schr\"odinger equation. Special attention is paid to the dependence of the electron emission pattern on the impact parameter $b$ of the ion with respect to the centre of the twisted wave front. In order to explore such a dependence, detailed calculations were carried out for the photoionization of the $1s$ ground and $2 \, p_y$ excited states of neutral hydrogen atoms. Based on these calculations, we argue that for relatively small impact parameters the electron angular distributions may be strongly affected by altering the position of the atom within the wave front. In contrast, if the atom is placed far from the front centre, the emission pattern of the electrons is independent on the impact parameter $b$ and resembles that observed in the photoionization by plane wave photons.
\end{abstract}


%
%
%
%
\section{Introduction}
\label{sec:introduction}

In the early 20th century, Einstein came up with his by now famous explanation for the photoelectric effect of UV--irradiated solid bodies \cite{einstein05}. Since then, the emission of electrons by light, also known as photoionization, has been studied extensively for various systems such as atoms \cite{cuellar91,mitsuke00,eichler07,blaga09}, atomic clusters \cite{misaizu92,gnodtke12} or molecules \cite{dill75,reid92,lux12}. Much attention, both in experiment and theory, has been paid to the total ionization cross sections and their dependence on photon energy and electronic configurations  \cite{pratt64,ichihara00,bauer08,chu09}. Apart from the fundamental interest, these studies play an important role for the precise description of ion charge state distributions in stellar as well as laboratory plasmas \cite{bautista98}.

\medskip

Besides the total rates, the \textit{angular} distributions of the photoelectrons have also been in the focus of intense research throughout the last decades \cite{johnson79, huang80, ichihara01, fritzsche08}. A large number of measurements were performed, for example, to explore the dependence of electron emission patterns on the polarization states of the incident photons \cite{cuellar91, mitsuke00, huang80}. These studies have revealed important information not only on the structure of many--electron systems but also on details of the electron--photon coupling. Owing to the recent advances in photo--optics, moreover, new possibilities arise to study the photoionization of atoms and molecules, in which the incoming light will possess not only a definite polarization but also a certain projection of the orbital angular momentum (OAM) onto its propagation direction \cite{allen92,beijersbergen93,terriza07}. These special states of light, also known as \textit{twisted photons}, can readily be generated in a 
number of ways, e.~g.~via computer--generated holograms, spiral phase plates, axicons or integrated ring resonators \cite{beijersbergen93,bazhenov92,heckenberg92,beijersbergen94,arlt00,cai12}. Moreover, by employing the Compton back-scattering of twisted optical photons off an ultra--relativistic electron beam, the production of high--energy photons with non-zero OAM may become feasible not only in the extreme ultraviolet (EUV) or x-ray domain but also in the region of GeV \cite{jentschura11-1,jentschura11-2}.

\medskip

There is a variety of light beams exhibiting OAM that may be used in photoionization studies. For example, the time--evolution of electron wavepackets during the ionization of hydrogen atoms by intense Laguerre--Gaussian pulses has been explored recently by Pic\'{o}n and co--workers \cite{picon10a,picon10b}. These theoretical investigations have provided the first fundamental insights into the electron dynamics in twisted light fields. The use of the Laguerre--Gaussian waves, however, imposes certain restrictions on the description of the photoionization process, since these waves satisfy the Helmholtz equation in the \textit{paraxial approximation} and, hence, their transverse spatial dimension has to be much smaller than the longitudinal one. In the present work, we extend the study of atomic ionization by twisted photons to the \textit{non--paraxial} regime by employing the so--called Bessel waves. These Bessel states are solutions of the full vector Helmholtz equation and carry not only orbital (OAM) but 
also spin angular momentum (SAM) which couple together to a total angular momentum (TAM) with a well--defined component along the direction of light propagation. During the recent years, Bessel beams of photons (as well as of charged particles) have been employed in a number of theoretical studies of various collision processes \cite{ivanov2011,iv-serbo2012} and of the photoexcitation of hydrogen atoms \cite{afanasev2013}.

\medskip

In order to perform a theoretical investigation of the angular distribution of electrons emitted in the course of atomic ionization by twisted photons, we will first recall the definition of the Bessel waves in terms of their vector potentials. By making use of these potentials, the orbital as well as the spin structure of the twisted states will be discussed in Sections \ref{subsec:orbital} and \ref{subsec:spin}, and the Poynting vector as well as the transversal intensity profile of the light beam will be evaluated later in Section~\ref{susec:intensity_profile}. Analysis of such a profile shows that the outcome of the photoionization study might depend on the position of the atom (or ion) within the wave front. In Section~\ref{subsec:geometry}, therefore, we specify the geometry under which the photoelectron emission is explored. We then proceed briefly in Section~\ref{subsec:ionization_plane_wave} with the well--known theoretical treatment of the conventional photoionization by plane wave photons. In 
analogy to this treatment, that is based on the first--order perturbation theory, we lay down a general procedure to investigate the photoeffect induced by twisted photons. The developed formalism is then applied to the ionization of the ground $1s$ as well as the excited 2$p_y$ states of neutral hydrogen atoms. Calculations, presented in Section~\ref{sec:results_and_discussion} clearly show the sensitivity of the photoelectron emission pattern to the impact parameter $b$ of the atom with regard to the twisted wave centre; the effect becomes most pronounced for small values of $b \lesssim 100-10000 \ {\rm a.u.}$ Finally, a short summary and an outlook are given in Section~\ref{sec:summary_and_outlook}.

\medskip

Atomic units (a.u.) are used throughout the paper, unless stated otherwise.

%
%
%
%
\section{Theoretical description of twisted photon states}
\label{sec:description}

In order to investigate the atomic photoionization by twisted (Bessel) photons, we first need to construct the corresponding states of light in terms of the 4--vector potential $A_\mu({\bf r},t)$. Such a potential should satisfy the wave equation 
\begin{equation}
\left (\Delta - \alpha^2 \frac{\partial^2}{\partial t^2} \right) A_\mu({\bf r},t)  = 0 \, ,
\end{equation}
which describes electromagnetic field configurations within the vacuum. Here $\Delta$ is the Laplace operator and $\alpha$ the electromagentic fine-structure constant. By imposing the Coulomb gauge condition and restricting ourselves to monochromatic states of light with a well--defined energy $\omega$, we find
\begin{equation}
   \label{eq:CoulombGauge}
   A_\mu({\bf r},t) = \left (0, \, {\bf A}({\bf r})\,\rme^{-\rmi \omega t} \right ) \,,\;\; {\rm div}{\bf A}({\bf r})=0 \, ,
\end{equation}
where the 3--vector ${\bf A}({\bf r})$ is a solution of the (free--space) Helmholtz equation
\begin{equation}
   \label{eq:VectorHelmholtzEquation}
   \left(\Delta + k^2 \right) {\bf A}({\bf r}) = 0 \, .
\end{equation}
In this expression $k = \alpha \, \omega$ denotes the wave number of the electromagnetic field \cite{rose55}. By making use of the linear momentum operator $\hat{\bf p}=-\rmi \nabla$, one can re--write Eq.~(\ref{eq:VectorHelmholtzEquation}) in the form
\begin{equation}
   \label{eq:sqaredMomentum}
   \hat{\bf p}^2 {\bf A}({\bf r}) = k^2 {\bf A}({\bf r}) \, ,
\end{equation}
which implies that the vector potential ${\bf A}({\bf r})$ is an eigenfunction of the squared momentum operator.

\medskip

Eqs.~(\ref{eq:CoulombGauge})--(\ref{eq:sqaredMomentum}) describe the vector potential of an electromagnetic field that propagates in free space and include both, the plane and the twisted wave solutions. For the latter case, however, some additional requirements have to be taken into account. That is, the potential ${\bf A}({\bf r})$ for the Bessel--type twisted light is expected to be an eigenfunction of the longitudinal momentum
\begin{equation}
   \label{eq:LongitudinalMomentum}
   \hat{p}_z \, {\bf A}({\bf r}) \equiv - \rmi \frac{\partial}{\partial z} {\bf A}({\bf r}) = k_z \, {\bf A}({\bf r})\,,
\end{equation}
and the $z$--component of the total angular momentum operator
\begin{equation}
   \label{eq:TotalAngularMomentum}
   \hat{J}_z \, {\bf A}({\bf r})= m_\gamma \, {\bf A}({\bf r}) \, ,
\end{equation}
where the operator $\hat{J}_z = \hat{L}_z + \hat{S}_z$ is given by the corresponding components of the orbital and spin angular momentum operators:
\begin{equation}
   \label{eq:orbital_spin_operators_explicit}
   \hat{L}_z=-\rmi \left( x \frac{\partial}{\partial y} - y \frac{\partial}{\partial x} \right ), \;\; \hat{S}_z=-\rmi
   \left(\begin{array}{ccc} 0 & 1 & 0 \\ -1 & 0 & 0 \\ 0 & 0 & 0 \end{array}\right) \, .
\end{equation}
From Eqs.~(\ref{eq:sqaredMomentum}) and (\ref{eq:LongitudinalMomentum}) it immediately follows, moreover, that the twisted photon is also characterized by a definite modulus of the transverse linear momentum
\begin{equation}
   \label{eq:TransversalMomentum}
   \left| {\bf k}_\perp \right| \equiv \varkappa = \sqrt{k^2 - k_z^2} \, .
\end{equation}
Having characterized the twisted photon state by Eqs.~(\ref{eq:VectorHelmholtzEquation})--(\ref{eq:TransversalMomentum}), we are ready now to find the explicit form of the vector potential ${\bf A}({\bf r})$. As will be shown in the next two sections, this requires an analysis of the orbital and spin properties of the Bessel solutions.

\subsection{Orbital structure of twisted states}
\label{subsec:orbital}

As seen from the discussion above, the vector potential ${\bf A}({\bf r})$ of twisted light is an eigenfunction of the $z$--component and of the square of the linear momentum operator $\hat{\bf p}$. In order to construct such solutions, we note first that the \textit{scalar} function
\begin{equation}
   \label{eq:ScalarFunction}
   \psi_{\varkappa, k_z, m_l}({\bf r}) \, = \,
   \sqrt{\frac{\varkappa}{2\pi}} \, J_{m_l}(\varkappa \, r_\perp) \, {\rme}^{{\rmi}m_l\varphi_r}\,{\rme}^{{\rmi}k_z z}
\end{equation}
also satisfies Eqs.~(\ref{eq:sqaredMomentum})--(\ref{eq:LongitudinalMomentum}). In this expression we used cylindrical coordinates $\left(r_\perp, \varphi_r, z\right) = \left(\sqrt{x^2 + y^2}, \arctan\left(y/x\right), z\right)$ and the Bessel function $J_{m}(x)$ of the first kind \cite{abramowitz70}. The function $\psi_{\varkappa, k_z, m_l}({\bf r})$ is normalized, moreover, as
\begin{equation}
   \label{eq:ScalarFunctionNormalization}
   \int \psi^\ast_{\varkappa^\prime, k^\prime_z, m^\prime_l }({\bf r})\,\psi_{\varkappa, k^{}_z, m^{}_l}({\bf r})\,{\rm d}^3 {\bf r}
   \, = \, 2\pi \delta(\varkappa-\varkappa^\prime) \, \delta(k^{}_z-k^\prime_z) \, \delta_{m^{}_l, m^\prime_l}\, ,
\end{equation}
and can be expressed as a superposition of plane waves
\begin{eqnarray}
   \label{eq:VectorPotentialTwistedPhoton}
   \psi_{\varkappa, k_z, m_l}({\bf r}) & = & \int a_{\varkappa, m_l}({\bf k}_\perp) \,\rme^{\rmi {\bf k}\cdot{\bf r}} \, \frac{{\rm d}^2 {\bf k}_\perp}{(2\pi)^2}\, \nonumber \\
   &\equiv& \int a_{\varkappa, m_l}({\bf k}_\perp) \,\rme^{\rmi \left({\bf k}_\perp {\bf r}_\perp + k_z z \right)} \, \frac{{\rm d}^2 {\bf k}_\perp}{(2\pi)^2} \, .
\end{eqnarray}
Here each (plane wave) component is weighted by the amplitude
\begin{equation}
   \label{eq:AmplitudesTwistedPhoton}
   a_{\varkappa, m_l}({\bf k}_\perp) = \sqrt{\frac{2 \pi}{\varkappa}} \, (-\rmi)^{m_l} \, \rme^{\rmi m_l \varphi_k} \, \delta(k_\perp - \varkappa)
\end{equation}
with $k_\perp = \left | {\bf k}_\perp \right |$ being the absolute value of the transverse momentum and $\varphi_k$ the azimuthal angle of the photon's wave vector ${\bf k}$, which, therefore, can be written as
\begin{equation}
   \label{eq:photon_wave_vector}
   {\bf k} \, = \, \left ( \begin{array}{c} k_\perp \cos\varphi_k \\ k_\perp \sin\varphi_k \\ k_z \end{array} \right ).
\end{equation}
We note that for fixed values of the transversal $\varkappa$ and longitudinal $k_z$ components of the linear momentum, all wave vectors ${\bf k}$ contributing to the integral (\ref{eq:VectorPotentialTwistedPhoton}) lie on a (momentum) cone with an opening angle $\theta_k = {\rm arctan}(\varkappa/k_z)$.

\medskip

The properties of the scalar function $\psi_{\varkappa, k_z, m_l}({\bf r})$ can be deduced directly from Eqs.~(\ref{eq:VectorPotentialTwistedPhoton}) and (\ref{eq:AmplitudesTwistedPhoton}). Namely, while the delta distribution $\delta(k_\perp - \varkappa)$ in the amplitudes~(\ref{eq:AmplitudesTwistedPhoton}) ensures that condition~(\ref{eq:TransversalMomentum}) is fulfilled, the factor of $\rme^{\rmi k_z z}$ in equation~(\ref{eq:VectorPotentialTwistedPhoton}) provides for requirement~(\ref{eq:LongitudinalMomentum}). Moreover, in the momentum space, where $\hat{L}_z=-\rmi \frac{\partial}{\partial \varphi_k}$, the amplitude $a_{\varkappa, m_l}({\bf k}_\perp)$ is an eigenfunction of the orbital angular momentum operator
\begin{equation}
   \label{orbitalInMomemtum}
   \hat{L}_z\,a_{\varkappa, m_l}({\bf k}_\perp) =  m_l \, a_{\varkappa, m_l}({\bf k}_\perp).
\end{equation}
By performing the Fourier transformation on both sides of Eq.~(\ref{orbitalInMomemtum}) in the 2-dimensional ${\bf k}_\perp$--space and by multiplying with $\rme^{\rmi k_z z}$, we find that a similar relation holds for the function $\psi_{\varkappa, k_z, m_l}({\bf r})$ in the coordinate space 
\begin{equation}
   \hat{L}_z\, \psi_{\varkappa, k_z, m_l}({\bf r})=m_l \, \psi_{\varkappa, k_z, m_l}({\bf r}).
\end{equation}
\subsection{Spin structure of twisted states}
\label{subsec:spin}

Beside the requirements (\ref{eq:sqaredMomentum}), (\ref{eq:LongitudinalMomentum}) and (\ref{eq:TransversalMomentum}) which define the orbital structure of the twisted Bessel light, the vector potential ${\bf A}({\bf r})$ has to be an eigenfunction of the ($z$--projection of the) \textit{total} angular momentum operator $\hat{J}_z = \hat{L}_z + \hat{S}_z$ (cf. Eq.~(\ref{eq:TotalAngularMomentum})). Therefore, to construct the ${\bf A}({\bf r})$ we have to specify also its spin properties. Along this line, let us first remind how one describes the spin--polarization properties of the standard \textit{plane wave} solutions of the Helmholtz equation (\ref{eq:VectorHelmholtzEquation}). Within the Coulomb gauge, the vector potential for such solutions reads as
\begin{equation}
   \label{eq:planeWave}
   {\bf A}^{\rm pl}({\bf r}) = {\bf e}_{{\bf k}, \Lambda}\,\rme^{\rmi {\bf k}\cdot{\bf r}}\,,\;\;{\bf e}_{{\bf k}, \Lambda} \cdot {\bf k}=0\, ,
\end{equation}
where the vector ${\bf e}_{{\bf k}, \Lambda}$ characterizes the photon with a certain \textit{helicity} $\Lambda = \pm 1$, i.e. the photon's spin projection onto its own momentum ${\bf k}$. In the general case, when ${\bf k}$ does not coincide with the quantization ($z$--) axis of the overall system and is given by Eq.~(\ref{eq:photon_wave_vector}), we can write the polarization vector ${\bf e}_{{\bf k}, \Lambda}$ in the form
\begin{equation}
   \label{eq:PolarizationVectorBesselBeam}
   {\bf e}_{{\bf k}, \Lambda} = \frac{-\Lambda}{\sqrt{2}} \left( \begin{array}{c}
   \cos\theta_k \cos\varphi_k - \rmi \Lambda \sin\varphi_k \\
   \cos\theta_k \sin\varphi_k + \rmi \Lambda \cos\varphi_k \\
   -\sin\theta_k \end{array} \right)\,.
\end{equation}
For practical applications, it is often convenient to present this vector as an expansion
\begin{equation}
   \label{eq:decomposition}
   {\bf e}_{{\bf k}, \Lambda} = \sum_{m_s = 0, \, \pm 1}
   c_{m_s} \, \rme^{-\rmi m_s \varphi_k} \, {\bm \eta}_{m_s} \, ,
\end{equation}
in the orthonormal basis $\{ {\bm \eta}_{m_s} \}_{m_s = 0, \, \pm 1}$ of the eigensolutions of the ($z$--component of) spin momentum operator:
\begin{equation}
   \label{eq:polarization_vectors_basis}
   \hat{S}_z {\bm \eta}_{m_s} \, = \, m_s \,{\bm \eta}_{m_s} , \;\;
   {\bm \eta}_{\pm 1} = \frac{\mp 1}{\sqrt{2}}
   \left( \begin{array}{c}  1 \\ \pm \rmi \\ 0 \end{array} \right), \;\;
   {\bm \eta}_{0}=\left( \begin{array}{c} 0 \\ 0 \\ 1 \end{array} \right) .
\end{equation}
Moreover, the expansion coefficients in Eq.~(\ref{eq:decomposition}) are given by:
\begin{equation}
   \label{eq:expansion_coefficients}
   c_{\pm 1}=\frac{1}{2} (1 \pm \Lambda \cos{\theta_k}), \;\;
   c_0= \frac{\Lambda}{\sqrt{2}} \, \sin{\theta_k}\,.
\end{equation}
Based on the decomposition (\ref{eq:decomposition}) one can easily prove that the polarization vector ${\bf e}_{{\bf k}, \Lambda}$ is an eigenfunction of the operator $\hat{J}_z$ corresponding to the eigenvalue \textit{zero}:
\begin{eqnarray}
   \label{eq:eigenfunction_of_Jz}
   \hat{J}_z \, {\bf e}_{{\bf k}, \Lambda} & \equiv & \left(\hat{L}_z + \hat{S}_z \right) \, \sum_{m_s = 0, \, \pm 1} c_{m_s} \, \rme^{-\rmi m_s \varphi_k} \, {\bm \eta}_{m_s} \nonumber\\
   & = & \sum_{m_s = 0, \, \pm 1} \left(-m_s + m_s \right) c_{m_s} \, \rme^{-\rmi m_s \varphi_k} \, {\bm \eta}_{m_s} = 0 \, .
\end{eqnarray}
Here we employed Eq.~(\ref{eq:polarization_vectors_basis}) and the trivial relation $\hat{L}_z \rme^{-\rmi m_s \varphi_k} = -m_s \rme^{-\rmi m_s \varphi_k}$ written in the momentum space, where $\hat L_z=-\rmi \frac{\partial}{\partial \varphi_k}$.

\subsection{Vector potential of twisted states and its properties}
\label{subsec:total_solutions}

Having discussed the orbital and the spin properties of the twisted--light solutions, we are ready now to write their explicit form. That is, the vector potential of the Bessel states can be expanded into plane waves as
\begin{eqnarray}
   \label{eq:twisted_light_vector_potential_explicit}
   {\bf A}_{\varkappa, k_z, m_\gamma, \Lambda}({\bf r}) & \equiv & {\bf A}^{\rm tw}({\bf r}) = \int a_{\varkappa, m_\gamma}({\bf k}_\perp) \, {\bf e}_{{\bf k}, \Lambda}\, \rme^{\rmi {\bf k}\cdot{\bf r}} \, \frac{{\rm d}^2 {\bf k}_\perp}{(2\pi)^2} \nonumber \\
   &=& \, \int a_{\varkappa, m_\gamma}({\bf k}_\perp) \, {\bf e}_{{\bf k}, \Lambda}\, \rme^{\rmi \left( {\bf k}_\perp {\bf r}_\perp + k_z z \right) } \, \frac{{\rm d}^2 {\bf k}_\perp}{(2\pi)^2}
\end{eqnarray}
As seen from this expression and Eq.~(\ref{eq:AmplitudesTwistedPhoton}), ${\bf A}_{\varkappa, k_z, m_\gamma, \Lambda}({\bf r})$ characterizes light with certain longitudinal ($k_z$) and transverse ($\varkappa$) components of the linear momentum, as it is requested by Eqs.~(\ref{eq:sqaredMomentum}), (\ref{eq:LongitudinalMomentum}) and (\ref{eq:TransversalMomentum}). Moreover, since the product of the $a_{\varkappa, m_\gamma}({\bf k}_\perp)$ and the polarization vector ${\bf e}_{{\bf k}, \Lambda}$ is an eigenfunction of the $z$--component of the total angular momentum operator
\begin{equation}
   \label{eq:orbital_function_polarization_vector_product}
   \hat{J}_z \, a_{\varkappa, m_\gamma}({\bf k}_\perp) \, {\bf e}_{{\bf k}, \Lambda} = m_\gamma \, a_{\varkappa, m_\gamma}({\bf k}_\perp)\, {\bf e}_{{\bf k}, \Lambda} \, ,
\end{equation}
as follows from Eqs.~(\ref{orbitalInMomemtum}) and (\ref{eq:eigenfunction_of_Jz}) written in the momentum space, the requirement (\ref{eq:TotalAngularMomentum}) is also fulfilled.

\medskip

Eq.~(\ref{eq:twisted_light_vector_potential_explicit}) provides the most general form of the vector potential of the twisted (Bessel) states and will be employed in the next sections to explore the angular distribution of the emitted photoelectrons. Before we start with such an analysis let us discuss first some basic properties of the twisted--light solutions. For example, by substituting the polarization vector (\ref{eq:decomposition}) into Eq.~(\ref{eq:twisted_light_vector_potential_explicit}) and performing the integration over the transverse photon momentum ${\bf k}_\perp$, we can write the vector potential ${\bf A}^{\rm tw}({\bf r})$ in terms of eigenfunctions (\ref{eq:polarization_vectors_basis}) of the spin operator $\hat{S}_z$ as
\begin{equation}
   \label{eq:BesselBeamFullForm}
   {\bf A}^{\rm tw}({\bf r}) = \sum_{m_s = 0, \, \pm 1}\, {\bm \eta}_{m_s} \, A^{\rm tw}_{m_s}({\bf r}) \, ,
\end{equation}
where the coefficients $A^{\rm tw}_{m_s}({\bf r})$ are given by
\begin{equation}
   \label{eq:expansion_coefficients_A}
   A^{\rm tw}_{m_s}({\bf r}) = \sqrt{\frac{\varkappa}{2\pi}} \, (-\rmi)^{m_s} \, c_{m_s} \, 
   J_{m_\gamma-m_s}(\varkappa \, r_\perp) \,
   \rme^{\rmi (m_\gamma-m_s) \varphi_r} \, \rme^{\rmi k_z z} \,.
\end{equation}
As seen from this expression, the potential ${\bf A}^{\rm tw}({\bf r})$ is a superposition of three terms with different projections of the orbital ($m_l = m_\gamma - m_s$) as well as spin ($m_s$) angular momentum. The projection of the total angular momentum for each term is then given by $m_l + m_s = m_\gamma$. Therefore, ${\bf A}^{\rm tw}({\bf r})$ possesses also a projection of the total angular momentum of $m_\gamma$ as required by constraint~(\ref{eq:TotalAngularMomentum}).

\medskip

One may further simplify Eqs.~(\ref{eq:BesselBeamFullForm})--(\ref{eq:expansion_coefficients_A}) if one assumes that the transverse momentum of the photon is much smaller comparing to its longitudinal momentum, $\varkappa << k_z$. Within such a \textit{paraxial approximation}, the summation in Eq.~(\ref{eq:BesselBeamFullForm}) is restricted to the single term $m_s = \Lambda$:
\begin{equation}
   \label{eq:BesselBeamFullFormParaxial}
   {\bf A}^{\rm tw}({\bf r}) = {\bm \eta}_{\Lambda} \, A^{\rm tw}_{\Lambda}({\bf r}) = {\bm \eta}_{\Lambda} \, \sqrt{\frac{\varkappa}{2\pi}} \, (-\rmi)^{\Lambda} \, c_{\Lambda} \, 
   J_{m_\gamma-\Lambda}(\varkappa \, r_\perp) \,
   \rme^{\rmi (m_\gamma-\Lambda) \varphi_r} \, \rme^{\rmi k_z z} \, .
\end{equation}
Eq.~(\ref{eq:BesselBeamFullFormParaxial}) indicates that the projections of the orbital and spin angular momenta onto the $z$-axis decouple within the paraxial approximation, i.~e.~ 
\begin{equation}
\hat{L}_z \, {\bf A}^{\rm tw}({\bf r}) = (m_\gamma - \Lambda) \, {\bf A}^{\rm tw}({\bf r}), \; \; \hat{S}_z \, {\bf A}^{\rm tw}({\bf r})= \Lambda \, {\bf A}^{\rm tw}({\bf r}) \, . 
\end{equation}
Moreover, in the limit $\varkappa \to 0$, where $J_{m_\gamma-\Lambda}(\varkappa \, r_\perp) \to \delta_{m_\gamma,\Lambda}$, Eq.~(\ref{eq:BesselBeamFullFormParaxial}) recovers the standard solution for a plane wave that propagates along the $z$--axis.

\begin{figure}[tb]
\includegraphics[width=0.5\textwidth]{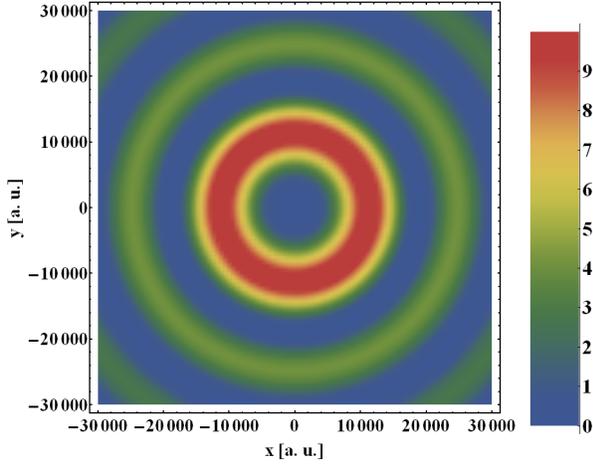}
\caption{\label{Fig:BesselBeamProfile} Transverse intensity profile of a Bessel wave (in arbitrary units) as given by Eq.~(\ref{eq:IntensityProfile}) for the parameters $k_z = 2.7 \cdot 10^{-2} \ {\rm a.u.}$, $\varkappa = 2.7 \cdot 10^{-4} \ {\rm a.u.}$, $\Lambda=+1$ and $m_\gamma = +3$. Here the typical ring structure of the intensity pattern with a well--defined (zero intensity) centre can be observed.}
\end{figure}
\section{Geometry of the photoionization process}
\label{sec:geometry}

\subsection{Poynting vector and intensity profile of twisted light}
\label{susec:intensity_profile}

To properly describe the geometry of the electron emission following the ionization of atoms by twisted waves, the spatial features of such waves have to be discussed. An appropiate quantity to characterize the spatial properties is given by the beam's (time--averaged) Poynting vector \cite{jackson62} that is defined via 
\begin{equation}
   \label{eq:PoyntingVector}
   {\bf P}({\bf r}) = g_P \, {\rm Re} \! \left[ \rmi \, {\bf A}^{\rm tw}({\bf r}) \times \left (\nabla \times {\bf A}^{\rm tw}({\bf r}) \right)^\ast \right ],
\end{equation}
where $g_P$ is a proportionality constant. By employing Eq.~(\ref{eq:expansion_coefficients}) as well as the decomposition (\ref{eq:BesselBeamFullForm}) and, furthermore, using (local) basis vectors of the cylindrical coordinate system
\begin{equation}
 {\bf e}_{r_\perp} = \left(\begin{array}{c} \cos\varphi_r \\ \sin\varphi_r \\ 0 \end{array} \right), \; \; {\bf e}_{\varphi_r} = \left(\begin{array}{c} -\sin\varphi_r \\ \cos\varphi_r \\ 0 \end{array} \right), \; \;  {\bf e}_{z} = \left(\begin{array}{c} 0 \\ 0 \\ 1 \end{array} \right),
\end{equation}
one may write the Poynting vector (cf.~Eq.~(13) in \cite{afanasev2013}) as
\begin{equation}
   \label{eq:PoyntingVectorCylindrical}
   {\bf P}({\bf r}) = P_{r_\perp}({\bf r}) \, {\bf e}_{r_\perp} + P_{\varphi_r}({\bf r}) \, {\bf e}_{\varphi_r} + P_{z}({\bf r}) \, {\bf e}_z,
\end{equation}
with
\begin{equation}
  \label{eq:PoyntingVectorR}
  P_{r_\perp}({\bf r}) = 0 \,,
\end{equation}
\begin{equation}
  \label{eq:PoyntingVectorPhi}
  P_{\varphi_r}({\bf r}) = g_P \frac{\varkappa^2}{2 \pi } J_{m_\gamma}(\varkappa \, r_\perp) \left ( c_{+1} J_{m_\gamma-1}(\varkappa \, r_\perp) + c_{-1} J_{m_\gamma+1}(\varkappa \, r_\perp) \right) 
\end{equation}
and
\begin{equation}
  \label{eq:PoyntingVectorZ}
  P_{z}({\bf r}) = g_P \, \Lambda \, \frac{ \varkappa k}{2 \pi} \left ( c_{+1}^2 J_{m_\gamma - 1}^2(\varkappa \, r_\perp) - c_{-1}^2 J_{m_\gamma + 1}^2(\varkappa \, r_\perp) \right ) \,.
\end{equation}
As seen from Eqs.~(\ref{eq:PoyntingVectorCylindrical})-(\ref{eq:PoyntingVectorZ}), no radial component of the Poynting vector is present and, therefore, we recover the well-known property that the Bessel beams are non--diffractive \cite{durnin87}. Furthermore, by taking the norm of the z-component of the Poynting vector, we derive the intensity profile of the Bessel beam within the plane that is perpendicular to the beam direction ($z$--axis) as 
\begin{equation}
   \label{eq:IntensityProfile}
   I_\perp({\bf r}) = \left |P_z({\bf r}) \right| \, .
\end{equation}
From Eqs.~(\ref{eq:PoyntingVectorZ})-(\ref{eq:IntensityProfile}) one can deduce that the (transverse) intensity profile $I_\perp({\bf r})$ has a pronounced radial structure that is given by the squared Bessel functions $J_{n}^2(\varkappa r_\perp)$ with $n = m_\gamma-1,m_\gamma+1$, but is independent, however, of $z$ and $\varphi_r$. In Fig.~\ref{Fig:BesselBeamProfile} we display, for example, such an intensity pattern for the parameters that will be used in our photoionization calculations below and are summarized in the second column of table~\ref{tab:BeamParameters}. In the centre of the figure, one can clearly see a \textit{zero--intensity} spot (also called vortex) which is typical for beams that carry orbital angular momentum \cite{allen92,beijersbergen93,terriza07}. This beam centre is surrounded by concentric rings of high and low intensity that alternate on a length scale of approximately $10^{4} \ {\rm a.u.}$ As will be shown in the next section, such a pronounced structure of the Bessel solutions 
in contrast to the plane waves leads to a much more complex geometry of the atomic photoionization process.

\begin{figure}[tb]
\includegraphics[width=0.5\textwidth]{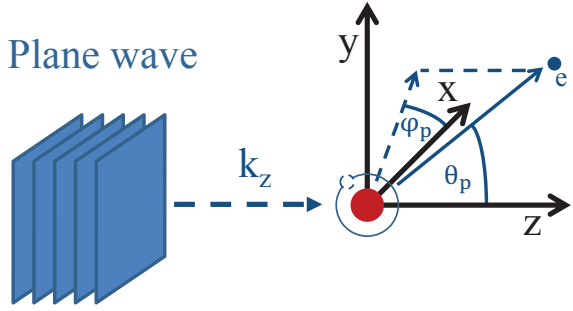}
\caption{\label{Fig:GeometryPlaneWave} Geometry of the atomic ionization by plane wave photons. The atomic nucleus is placed at the origin of the coordinate system, whose $z$--axis is chosen along the propagation direction of the photon beam. For the $K$--shell ionization, moreover, the direction of the $x$-- and $y$--axes are not predefined by the photon--atom system and can be chosen arbitrarily. In contrast, for an atomic state which has a pronounced structure of its electron cloud, such as the $p_y$--level \cite{bransden83}, one can utilize this structure to define the $y$--axis.}
\end{figure}
\subsection{Characterization of the electron emission}
\label{subsec:geometry}

In order to explore the atomic photoionization, we shall first agree about the geometry and coordinates under which the electron emission is observed. Before starting such a discussion for the twisted waves, with their pronounced structure (cf.~Fig.~\ref{Fig:BesselBeamProfile}), let us briefly remind how one characterizes the photoelectrons if the incident light is a plane wave. In this case, it is natural to adopt the position of the atom as the origin of a coordinate system, whose $z$--axis (quantization axis) is taken along the propagation direction of the waves (cf.~Fig.~\ref{Fig:GeometryPlaneWave}). In order to choose properly the direction of the $x$-- and $y$--axes, additional information is required about the spin state of the incident light or the atomic sublevel population. For the $K$--shell ionization by circularly polarized (or unpolarized) photons, for example, the overall system possesses an axial symmetry and, hence, the $x$-- and $y$--axes can be adopted arbitrarily. A single polar angle $\
theta_p$ is needed in this case to characterize the emitted electron. If, in contrast, the atom was \textit{prepared} before the ionization in the excited $p_y$--state, in which the electron density is not spherically symmetric but oriented along some preferred direction, we can employ this direction to define the $y$--axis.

\begin{figure}[tb]
\includegraphics[width=0.5\textwidth]{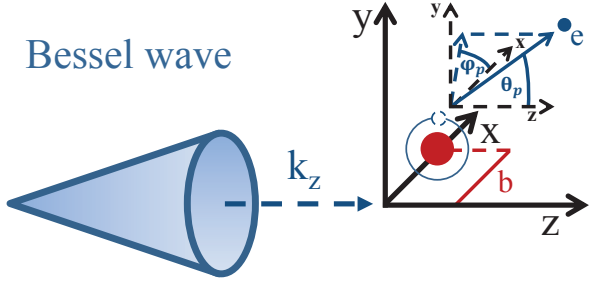}
\caption{\label{Fig:GeometryBesselBeam}
Geometry of the atomic ionization by Bessel photons. The $z$--axis is chosen along the beam's direction of propagation, while the $x$--$z$ (reaction) plane is determined by the position of the atom and the zero--intensity line of the Bessel wave (as indicated by the centre of the cone).}
\end{figure}

\medskip

As mentioned already, the Bessel waves---in contrast to plane wave photons---have a pronounced spatial profile which implies a more complex geometry of the ionization process. Apart from the z--axis which is chosen similar as before along the direction of the wave propagation, the $x$--$z$ (reaction) plane has to be introduced. The plane passes through the position of the atom and the zero--intensity line as shown on Fig.~\ref{Fig:GeometryBesselBeam}. Therefore, except for the case when the atom is located just in the centre of the wave front, the emission of the electron following the ionization by twisted light is described by two angles $\hat{n}_p = \left( \theta_p \, , \varphi_p \right)$. The dependence of the differential (ionization) cross--section on these two angles will be studied below for different distances between the atom and the zero--intensity line as characterized by the \textit{impact parameter} $b$.

\section{Angular distribution of the photoelectrons}
\label{sec:angular_distribution}

Having discussed the vector potential of the twisted light and the geometry of the ionization process, we are ready now to study the angular distribution of the emitted electrons. Again, we start this investigation from the simple case of incident plane waves. Then, by analogy with the well--established plane wave treatment, we will lay down a general formalism for the description of the atomic ionization by Bessel beams.

\subsection{Atomic ionization by plane wave photons}
\label{subsec:ionization_plane_wave}

Not much has to be said about the theoretical description of the ionization of one--electron ions by plane wave photons. Within the first--order perturbation theory any analysis of this process can be traced back to the evaluation of the matrix element
\begin{equation}
   \label{eq:PlaneWaveMatrixElement}
   M_{fi}^{\rm pl}(\theta_p,\varphi_p) = \int \psi_f^\ast({\bf r})
   \left ({\bf A}^{\rm pl}({\bf r}) \cdot \hat{\bf p} \right )
   \psi_i({\bf r}) {\rm d}^3 \textbf{r} \, ,
\end{equation}
which describes the transition between initial (bound) and final (continuum) electron states under absorption of light whose properties are characterized by the vector potential (\ref{eq:planeWave}). In Eq.~(\ref{eq:PlaneWaveMatrixElement}), moreover, $\hat{\bf p}$ is the electron momentum operator and the dependence of the transition amplitude on the electron emission angles $(\theta_p,\varphi_p)$ results from the continuum wavefunction.

\medskip

As seen from Eq.~(\ref{eq:PlaneWaveMatrixElement}), the calculation of the matrix element $M_{fi}^{\rm pl}$ requires the knowledge of the bound-- as well as continuum--state wavefunctions. Since in the present work we aim to explore the (photoelectron) emission patterns for different \textit{shapes} of the electron cloud in the initial atom, the $\psi_i({\bf r})$ will be taken below as (i) the well--known Schr\"odinger solution
\begin{equation}
   \label{eq:HydrogenlikeWavefunctionPositionSpace}
   \psi_i({\bf r}) \equiv \psi_{n,l,m}({\bf r}) = R_{n,l}(r) \, Y_{l,m}(\theta_r,\varphi_r) \, ,
\end{equation}
characterized by the principal quantum number $n$, the orbital angular momentum $l$ and its projection $m$ onto the $z$--axis, and (ii) the wavefunction of the 2$p_y$ level
\begin{equation}
   \label{eq:pylevel}
    \psi_i({\bf r}) \equiv \psi_{p_y}({\bf r}) = \frac{1}{2 \rmi} \left ( \psi_{2, 1, +1}({\bf r}) + \psi_{2, 1, -1}({\bf r}) \right ) \, .
\end{equation}
While the function $\psi_{p_y}({\bf r})$ characterizes a state whose electron density is oriented along the $y$--axis, the density $\rho_{n,l,m}({\bf r}) = \left| \psi_{n,l,m}({\bf r}) \right|^2$ of the state~(\ref{eq:HydrogenlikeWavefunctionPositionSpace}) is axially symmetric around the $z$--axis \cite{bransden83}.

\medskip

The final--state electron is described in our investigation by a plane wave
\begin{equation}
   \label{eq:PlaneWaveElectron}
   \psi_f({\bf r}) \equiv \psi_{{\bf p}}({\bf r}) = \rme^{\rmi {\bf p}\cdot{\bf r}} \, ,
\end{equation}
which constitutes the well--known Born approximation \cite{bethe57}. Within such an approximation that is valid for (electron) kinetic energies much larger than the ionization threshold, one neglects the electron--ion attraction after the ionization process. The great advantage of the approximation (\ref{eq:PlaneWaveElectron}) is that it allows for a simple analytical evaluation of the transition amplitude $M_{fi}^{\rm pl}(\theta_p,\varphi_p)$. By inserting, for example, wavefunctions (\ref{eq:HydrogenlikeWavefunctionPositionSpace}) and (\ref{eq:PlaneWaveElectron}) into Eq.~(\ref{eq:PlaneWaveMatrixElement}), we find
\begin{eqnarray}
   \label{eq:FinalPlaneWaveMatrixElement}
   M_{fi}^{pl}(\theta_p,\varphi_p) & \equiv &  M_{\, n, l, m}^{pl}(\theta_p,\varphi_p) =
   ({\bm \eta}_\Lambda \cdot {\bf p})
   \int \rme^{-\rmi({\bf p}-{\bf k}){\bf r}} \, \psi_{n,l,m}({\bf r}) {\rm d}^3 {\bf r} \nonumber \\
   & = & (2\pi)^{3/2} ({\bm \eta}_\Lambda \cdot {\bf p}) \, \tilde{\psi}_{n, l, m}({\bf q}) \, ,
\end{eqnarray}
where ${\bf q} = {\bf p} - {\bf k}$ and the explicit form of the Fourier transform $\tilde{\psi}_{n, l, m}({\bf q})$ is given in the Appendix A. With the help of this expression, moreover, one can immediately construct the ionization amplitude for the $p_y$ state (\ref{eq:pylevel}) as
\begin{equation}
   \label{eq:Final_py_level_amplitude}
   M_{fi}^{pl}(\theta_p,\varphi_p) \equiv  M_{p_y}^{pl}(\theta_p,\varphi_p) =
   \frac{1}{2i} \left( M_{\, 2, 1, +1}^{pl}(\theta_p,\varphi_p) + M_{\, 2, 1, -1}^{pl}(\theta_p,\varphi_p) \right) \, .
\end{equation}

In Section~\ref{sec:results_and_discussion} we will employ the matrix elements (\ref{eq:FinalPlaneWaveMatrixElement})--(\ref{eq:Final_py_level_amplitude}) in order to calculate the angular distribution of the emitted photoelectrons. Such a distribution is just given by the square of the amplitude
\begin{equation}
   \label{eq:AngularDistributionPlaneWave}
   W^{\rm pl}(\theta_p,\varphi_p) = N^{\rm pl} \left| M_{fi}^{pl}(\theta_p,\varphi_p) \right|^2 \, ,
\end{equation}
where $N^{\rm pl}$ is a normalization factor whose choice will be discussed later.

\subsection{Atomic ionization by twisted photons}
\label{subsec:ionization_twisted_wave}

Similar to the plane wave case from above, we start the treatment of the atomic ionization by twisted Bessel photons with the discussion of the transition amplitude
\begin{equation}
   \label{eq:TwistedPhotonMatrixElement}
   M_{fi}^{\rm tw}(\theta_p, \varphi_p) = \int \psi_f^\ast({\bf r}) \left ({\bf A}^{\rm tw}({\bf r})\cdot \hat{\bf p} \right ) \rme^{-\rmi {\bf b}\cdot\hat{\bf p}} \, \psi_i({\bf r}) \, {\rm d}^3 \textbf{r} \, .
\end{equation}
Again, $\psi_i$ and $\psi_f$ denote here the initial (bound) and final (continuum) electron wavefunctions, given by Eqs.~(\ref{eq:HydrogenlikeWavefunctionPositionSpace})--(\ref{eq:pylevel}) and (\ref{eq:PlaneWaveElectron}), correspondingly. Moreover, the vector potential of the incident twisted light is defined by expression (\ref{eq:twisted_light_vector_potential_explicit}), and the additional operator $\rme^{-\rmi {\bf b} \hat{\bf p}}$ translates the bound--electron wavefunction from the centre of the beam
\begin{equation}
   \label{eq:translation_operator}
   \rme^{-\rmi {\bf b}\cdot\hat{\bf p}} \psi_i({\bf r}) = \psi_i({\bf r}-{\bf b}) \, ,
\end{equation}
thus incorporating the impact parameter of the atom with respect to the zero--intensity line (cf.~Fig.~\ref{Fig:GeometryBesselBeam}).

%
%
\Table{\label{tab:BeamParameters} Parameters of the twisted Bessel light used in the photoionization calculation.}
\br
light parameter & first scenario & second scenario & third scenario \\
\mr
photon energy $E_\gamma$ & $100 \ {\rm eV}$ & $100 \ {\rm eV}$ & $100 \ {\rm eV}$ \\
longitudinal momentum $k_z$ & $2.68 \cdot 10^{-2} \ {\rm a.u.}$ & $2.67 \cdot 10^{-2} \ {\rm a.u.}$ & $1.89 \cdot 10^{-2} \ {\rm a.u.}$ \\
transversal momentum $\varkappa$ & $2.68 \cdot 10^{-4} \ {\rm a.u.}$ & $2.67 \cdot 10^{-3} \ {\rm a.u.}$ & $1.89 \cdot 10^{-2} \ {\rm a.u.}$ \\
opening angle $\theta_k$ & $0.57^\circ$ & $5.71^\circ$ & $45^\circ$ \\
ratio $s=\varkappa/k_z$ & $0.01$ & $0.1$ & $1$ \\
helicity $\Lambda$ & $+1$ & $+1$ & $+1$ \\
$z$--component of TAM $m_\gamma$ & $+3$ & $+3$ & $+3$ \\
\br
\endTable

\medskip

As for the evaluation of the plane wave matrix element (\ref{eq:FinalPlaneWaveMatrixElement}), we can further simplify  Eq.~(\ref{eq:TwistedPhotonMatrixElement}) if we employ the Fourier transformation of the bound--state wavefunction (\ref{eq:HydrogenlikeWavefunctionPositionSpace}):
\begin{eqnarray}
   \label{eq:twisted_amplitude_final}
   M_{fi}^{\rm tw}(\theta_p,\varphi_p) & \equiv & M_{n, l, m}^{\rm tw}(\theta_p,\varphi_p) \nonumber \\
   & = & (-\rmi)^{m_\gamma} \sqrt{\varkappa} \int_0^{2 \pi} \rme^{\rmi m_\gamma \varphi_k}
   \rme^{- \rmi {\bf b}\cdot{\bf q}} \, ({\bf e}_{{\bf k}, \Lambda} \cdot
   {\bf p}) \, \tilde{\psi}_{n, l, m}({\bf q}) \, {\rm d} \varphi_k \, .
\end{eqnarray}
By making use of the residue theorem from complex variable theory \cite{rudin87}, the integral over the azimuthal angle $\varphi_k$ in Eq.~(\ref{eq:twisted_amplitude_final}) can be calculated analytically. However, since the resulting expression is rather lengthy, we will not present it here and refer the interested reader to the Appendix B.

\medskip

In a full analogy with Eq.~(\ref{eq:AngularDistributionPlaneWave}), the square of the transition amplitude (\ref{eq:TwistedPhotonMatrixElement}) describes---up to some normalization factor and within the non--relativistic framework---the angular distribution of the electrons emitted due to atomic ionization by twisted Bessel light. This angular distribution is obtained for the case when the vector potential ${\bf A}^{\rm tw}({\bf r})$ describes the most general solutions of the vector Helmholtz equation that have a well--define projection of the total angular momentum onto the $z$--direction. Our analysis goes, therefore, beyond the paraxial approximation in contrast to previous photoionization studies \cite{picon10a,picon10b}. In the next section we apply the derived formalism to investigate the electron emission from different bound states of a neutral hydrogen atom.

\begin{figure}[tb]
\includegraphics[width=\textwidth,clip=true]{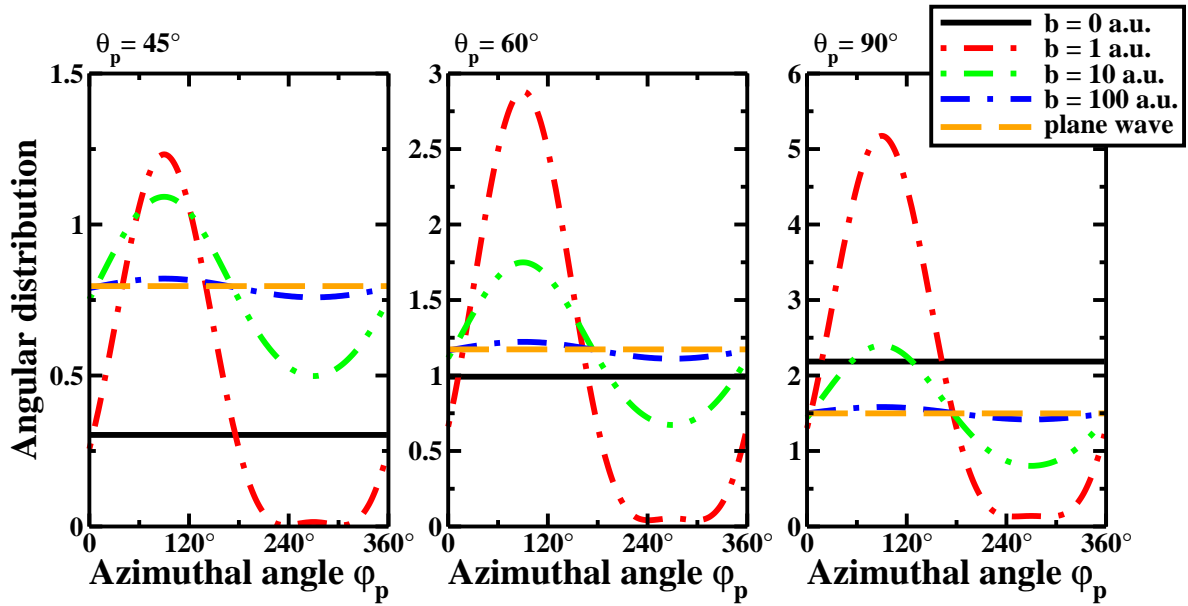}
\caption{\label{Fig:AngularDistributionGroundState}
Angular distribution of electrons emitted in the $K$--shell ionization of a neutral hydrogen atom by twisted light. Results are presented for different impact parameters $b$, that characterize the position of the atom in the wavefront (cf.~Fig.~\ref{Fig:GeometryBesselBeam}), and three polar angles $\theta_p = 45^\circ$ (left panel), $\theta_p = 60^\circ$ (middle panel) and $\theta_p = 90^\circ$ of the (continuum) electron. For comparison, the emission pattern predicted for the incoming plane wave photons with helicity $\Lambda = + 1$ is depicted by the dashed line.}
\end{figure}
%
%
%

%
%
\section{Results and discussion}
\label{sec:results_and_discussion}

In the previous section, we have laid out a general formalism for the description of the angular distribution of electrons emitted in the ionization of hydrogen--like systems by twisted Bessel light. While, of course, this theory can be applied to the photoemission from any (one--electron) bound state, we first focus our analysis on the $K$--shell ionization of a neutral hydrogen atom. Furthermore, we have to agree on the properties of the twisted photon beam. In our calculations, we will describe this beam by either of three different sets of beam parameters that are summarized in table~\ref{tab:BeamParameters}. For all scenarios described in table~\ref{tab:BeamParameters}, the energy of the incident light is chosen to be $E_\gamma = \omega \equiv$ 100~eV that---being well above the $1s$ ionization threshold---ensures the validity of the Born approximation (\ref{eq:PlaneWaveElectron}). 

\medskip

Having defined the prerequisites, let us now discuss the ionization process for the scenario that is characterized by the second column of table~\ref{tab:BeamParameters}. Here the ratio of transverse to longitudinal components of the photon's linear momentum, $s = \varkappa / k_z$, amounts to $s = 0.01$ which corresponds to the paraxial approximation~(\ref{eq:BesselBeamFullFormParaxial}). The electron emission pattern evaluated for such a set of parameters is displayed in Fig.~\ref{Fig:AngularDistributionGroundState} as a function of the azimuthal angle $\varphi_p$. Calculations have been performed for impact parameters in the range 0~$\le b \le$~100~a.u. and for three polar angles $\theta_p=45^\circ$ (left panel), $\theta_p=60^\circ$ (middle panel) and $\theta_p=90^\circ$ (right panel). Our predictions are compared, moreover, with the result obtained for the ground--state ionization by plane wave photons with helicity $\Lambda = +1$. For both, the plane wave-- and the twisted--light scenarios the angular 
distributions are normalized as
\begin{equation}
   \label{eq:angular_distribution_normalization}
   \int W^{{\rm tw}, \, {\rm pl}}(\theta_p, \varphi_p) \, {\rm d}\Omega_p = 4 \pi \, ,
\end{equation}
thus allowing us to study how the \textit{shape} of the electron emission patterns is changed for different impact parameters.

\begin{figure}[tb]
\includegraphics[width=\textwidth,clip=true]{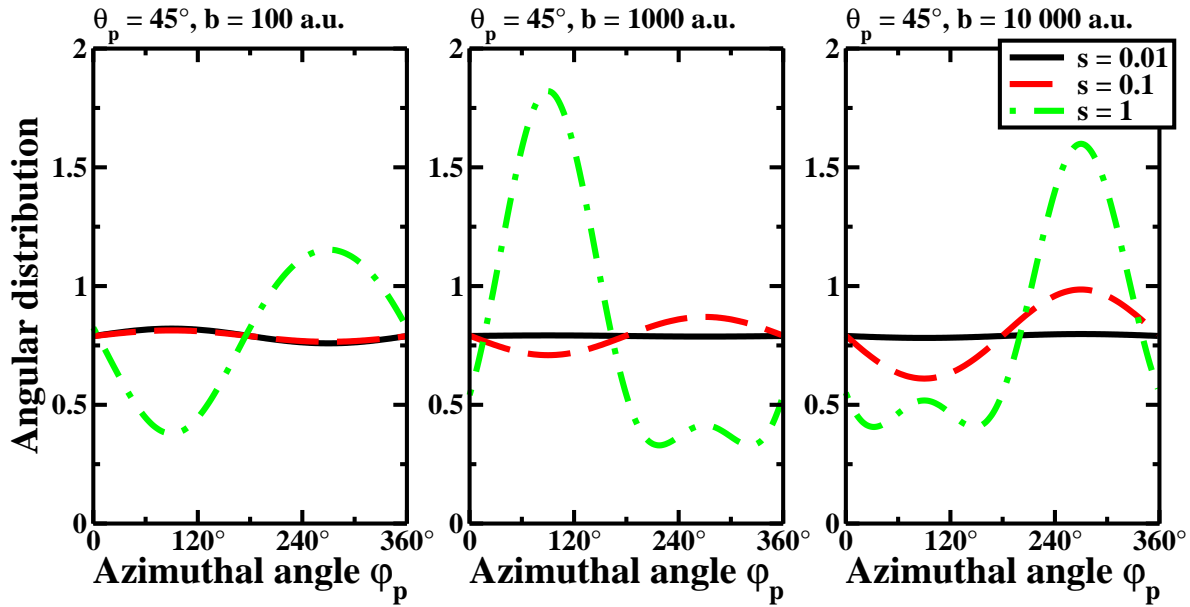}
\caption{\label{Fig:AngularDistributionGroundStateTransversalMomentum}
Angular distribution of electrons emitted in the $K$--shell ionization of a neutral hydrogen atom by twisted light. Results are presented for three sets of parameters of the incident light with different values for the ratio $s=\varkappa/k_z$ of the transversal to longitudinal momentum as given in table~\ref{tab:BeamParameters}. Moreover, the electron polar angle is taken as $\theta_p=45^\circ$, whereas the impact parameters are $b = 100 \ {\rm a.u.}$ (left panel), $b = 1000 \ {\rm a.u.}$ (middle panel) and $b = 10000 \ {\rm a.u.}$ (right panel).}
\end{figure}

\medskip

As seen from Fig.~\ref{Fig:AngularDistributionGroundState}, the angular distribution $W^{{\rm tw}}(\theta_p, \varphi_p)$ is isotropic if the atom is placed in the centre of the wavefront. This is well expected since the system of ``atom in the $1s$ state + Bessel beam'' possesses for $b = 0$ cylindrical symmetry about the $z$--axis and, hence, the photoionization probability is independent on the azimuthal angle. Such a symmetry is broken if one shifts the atom position from the zero--intensity centre. A remarkable anisotropy of the electron emission pattern can be observed, therefore, for impact parameters in the range 0~$ < b \lesssim$~100~a.u. However, if the distance between the atom and the wavefront centre becomes very large, $b > 100$~a.u., the angular distribution $W^{{\rm tw}}(\theta_p, \varphi_p)$ converges to the one that results from the plane wave photoionization. Such an impact--parameter--behaviour can be easily understood if one compares the characteristic scales of the atomic target and the 
Bessel beam. Namely, while the intensity profile of the light front changes notably on a length scale of $10^4 \ {\rm a.u.}$ (see Fig.~\ref{Fig:BesselBeamProfile}), the ground--state electron is confined to a volume characterized by a linear size of a few atomic units. Therefore, the electron does not ``see'' the large--scale (radial) variations of the light intensity profile and its emission can be affected only by the \textit{phase structure} of the Bessel photons as defined by the terms $\rme^{\rmi m \varphi_r}$ in Eqs.~(\ref{eq:BesselBeamFullForm})--(\ref{eq:expansion_coefficients_A}). Obviously, the change of this phase structure on the scale of an atomic Bohr radius can be remarkable only for relatively small parameters $b$. In contrast, far away from the light--wave axis the $\rme^{\rmi m \varphi_r}$ term varies slowly over the extension of the atom and the electron cloud is exposed to an (almost) constant phase. For $b > 100$~a.u., therefore, the ionization by twisted light resembles the one by plane 
wave photons as one may observe in Fig.~\ref{Fig:AngularDistributionGroundState}.

\begin{figure}[tb]
\includegraphics[width=\textwidth,clip=true]{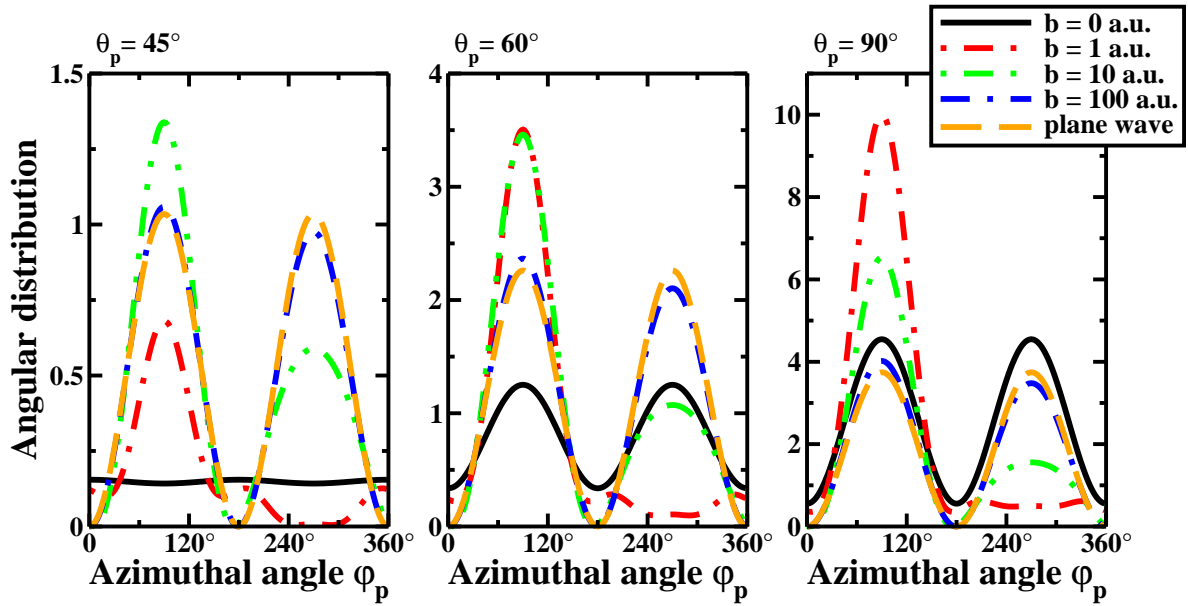}
\caption{\label{Fig:AngularDistributionpy} Same as in figure \ref{Fig:AngularDistributionGroundState} but for the $p_y$--state of hydrogen atoms.}
\end{figure}

\medskip

Until now we have discussed the impact--parameter dependence of the angular distribution $W^{{\rm tw}}(\theta_p, \varphi_p)$ for the parameters of the incident light given in the second column of table~\ref{tab:BeamParameters}. In particular, we have restricted our analysis to the case where the transversal component $\varkappa$ of the photon's linear momentum is negligible with respect to the longitudinal component $k_z$. In order to understand how the electron emission pattern changes if one departs from such a \textit{paraxial} approximation, calculations have been performed for the same photon energy and helicity as before but for three different values of the parameter $s = \varkappa/k_z$: $s = 0.01$, $s=0.1$ and $s=1$ (cf.~table~\ref{tab:BeamParameters}). While the first of these values obviously corresponds to the paraxial limit, the last one describes the general (non--paraxial) regime. In Fig.~\ref{Fig:AngularDistributionGroundStateTransversalMomentum} we display the angular distribution $W^{{\rm tw}
}(\theta_p, \varphi_p)$ that has been evaluated for these $s$ values as well as for the impact parameters $b$~=~100, 1000 and 10000 ${\rm a.u.}$ As seen from the figure, the electron emission pattern appears to be very sensitive to the variation of $s$. That is, while $W^{{\rm tw}}(\theta_p, \varphi_p)$ is almost isotropic in the paraxial regime ($s = 0.01$), it shows a remarkable $\varphi_p$--dependence if the transversal component of the photon linear momentum is comparable to the longitudinal one ($s \sim 1$). In order to understand this sensitivity to the component ratio $s$ one has to revisit Eq.~(\ref{eq:IntensityProfile}), which describes the intensity profile of the twisted wavefront. This equation suggests that the length scale of the intensity oscillations depends---via the squares of the Bessel functions---on the value of the transversal momentum. For large $\varkappa$, and, hence, $s \sim 1$, the period of oscillations is of the length scale of $100 \ {\rm a.u.}$ and becomes comparable with the 
size of the atomic ground state. The electron wavefunction may ``feel'', therefore, the features of the intensity distribution of the photon beam and this is reflected in the angular distribution as shown in Fig.~\ref{Fig:AngularDistributionGroundStateTransversalMomentum}.

\medskip

The $1s$--ground state of the hydrogen atom, whose ionization has been discussed so far, possesses spherical symmetry. The $\varphi_p$--dependence of the photoelectron emission pattern was caused, therefore, by the displacement of the target atom with respect to the zero--intensity centre of the wavefront. Apart from such a displacement, one can break the cylindrical symmetry of the overall system ``target + incident light'' if one prepares the initial atom in some \textit{oriented} state. In the present work, for example, we investigated the ionization of the hydrogenic 2$p_y$ state (\ref{eq:pylevel}) by twisted light that is characterized by the parameters given in the second column of table~\ref{tab:BeamParameters}. The angular distribution $W^{{\rm tw}}(\theta_p \, , \, \varphi_p)$, computed for these parameters and for different positions of the atom within the wavefront, is displayed in Fig.~\ref{Fig:AngularDistributionpy} as a function of the azimuthal angle $\varphi_p$ . As expected from the 
discussion of the symmetry properties of the 2$p_y$ state, the $W^{{\rm tw}}(\theta_p, \varphi_p)$ is remarkably anisotropic even at \textit{zero} impact parameter. With the increase of $b$, the shape of the angular distribution varies significantly until it resembles---for $b > 100$~a.~u.~---the emission pattern of the electrons emitted in the atomic ionization by plane wave photons (cf.~Fig.~\ref{Fig:AngularDistributionpy}). This impact--parameter--behaviour can be explained by the same reasoning as given for the $K$--shell ionization.

\medskip

As seen from Figs.~\ref{Fig:AngularDistributionGroundState}--\ref{Fig:AngularDistributionpy}, a pronounced $\varphi_p$--behaviour of the electron angular distribution $W^{{\rm tw}}(\theta_p, \varphi_p)$ that is attributed to the phase structure (paraxial regime) or to the intensity profile (non--paraxial regime) of the twisted waves can be observed for impact parameters in the range $0 < b < 100$~a.u. and $0 < b < 100000$~a.u., respectively. The experimental measurement of the predicted effects would require, therefore, an operational control of the parameter $b$. Even though the practical realization of such a control is a very complicated task, the recent advances in atomic trap technologies suggest that it might become feasible in the near future. For example, it is currently feasible to trap an atom with a spatial uncertainty of several ${\rm nm}$ \cite{wilson03,eschner03,murphy09}, which is just of the order where the characteristic photoionization features of the Bessel beams become visible (cf.~Fig.~\
ref{Fig:AngularDistributionGroundStateTransversalMomentum}).

%
%
\section{Summary and outlook}
\label{sec:summary_and_outlook}

In this work we have performed a theoretical analysis of the atomic ionization by twisted (Bessel) photons. By making use of the non--relativistic Schr\"odinger theory and the first--order perturbative approach, we laid down a general formalism for the description of the photoelectron angular distribution. While this theory can be employed to analyze the ionization of an arbitrary hydrogenic level, detailed calculations have been performed for the electron emission from the $1s$ ground and 2$p_y$ excited states. Our results indicate that the emission pattern is very sensitive to the position of the target atom within the wave as characterized by the impact parameter $b$ with respect to the zero--intensity wavefront centre. For relatively small impact parameters, $b \lesssim$~100~a.u.~for the paraxial and $b \lesssim$~10000~a.u.~for the non--paraxial light beams, the electron angular distribution reflects the phase structure and the intensity profile of the Bessel solutions, respectively. In contrast, if the 
atom is displaced far away from the beam centre, the electron emission remains almost unaffected by the intensity and phase structure of the twisted beam and resembles the one that is observed for the ionization by plane waves.

\medskip

Our present study provides a theoretical basis for the analysis of the atomic photoeffect in the most general case, i.e.~it is not restricted to the paraxial regime of the (twisted) light propagation. It complements, therefore, previous photoionization studies \cite{picon10a,picon10b} and opens up a way for further and deeper investigations of the ionization properties. In particular, by making use of the derived expressions we aim to explore the ionization of Rydberg atoms and to analyze the orbital angular momentum states of the emitted photoelectrons. This will help to clarify the question whether and how the photoeffect can be employed for the production of twisted electrons. Moreover, the developed analytical approach can be extended to study other fundamental processes, such as, for example, elastic scattering of twisted electrons by atoms and ions. Theoretical analysis of this process is currently underway and will be presented in a forthcoming paper.

%
%
\section*{Acknowledgements}

O.~M. and A.~S. acknowledge support from the Helmholtz Gemeinschaft and GSI (Nachwuchsgruppe VH-NG-421). A.~G.~H. acknowledges the support from the GSI Helmholtzzentrum and the University of Heidelberg. V.~G.~S. is supported by the Russian Foundation for Basic Research via grants 13--02--00695 and NSh--3802.2012.2.

%
%
%
%

\appendix

\section{Fourier transform of the bound--state wavefunction}
\label{app:PlaneWaveMatrixElements}

As seen from Eqs.~(\ref{eq:PlaneWaveMatrixElement}) and (\ref{eq:FinalPlaneWaveMatrixElement}), the evaluation of the amplitude for the bound--free electron transition under the absorption of the plane wave photons can be traced back to the Fourier transform of the wavefunction (\ref{eq:HydrogenlikeWavefunctionPositionSpace}). Within the non--relativistic framework, this transform reads as \cite{podolsky29, drake96}
\begin{eqnarray}
   \label{eq:WaveFunctionMomentumSpace}
   \tilde{\psi}_{n, l, m}({\bf p}) & = & 2^{2l+2} \, l! \, (-\rmi)^l \, \ n^{-l-2} \, \sqrt{\frac{2 \, (n-l-1)!}{\pi Z^3 (n+l)!}} \left(\frac{p}{Z} \right)^l \frac{Z^{2l+4}}{({p}^2 + \delta^2)^{l+2}} \nonumber \\
   & \times & C_{n-l-1}^{l+1}\!\left(\frac{{p}^2 - \delta^2}{{p}^2 + \delta^2} \right) Y_{l, m}(\theta_{p},\varphi_{{p}}),
\end{eqnarray}
where $\delta = Z / n$ with the nuclear charge $Z$ and $C_{n-l-1}^{l+1}(x)$ is a so--called Gegenbauer polynomial \cite{abramowitz70}.

\section{Matrix elements for photoionization by twisted photons}
\label{sec:TwistedMatrixElements}

Similar to the ionization by plane wave photons, any analysis of the electron emission induced by the incident twisted light requires the knowledge of the transition amplitude (\ref{eq:TwistedPhotonMatrixElement}). As discussed in Section~\ref{subsec:ionization_twisted_wave}, this amplitude can be written, upon the Fourier transformation of the bound--state wavefunction, in the form (\ref{eq:twisted_amplitude_final}). In order to proceed further, we insert the explicit expression (\ref{eq:WaveFunctionMomentumSpace}) of $\tilde{\psi}_{n, l, m}({\bf q})$ into Eq.~(\ref{eq:twisted_amplitude_final}) and find
\begin{eqnarray}
  \label{eq:OriginalIntegral}
  M_{fi}^{\rm tw}(\theta_p,\varphi_p) & = & c_{\rm tw} \int_0^{2 \pi} \rme^{\rmi m_\gamma \varphi_k} \, \rme^{-\rmi {\bf b}\cdot{\bf q}} \, q^l \frac{({\bf e}_{k, \Lambda} \cdot {\bf p})}{(q^2+\delta^2)^{l+2}} \nonumber\\ 
  & \times & C_{n-l-1}^{l+1} \left(\frac{q^2-\delta^2}{q^2+\delta^2}\right) \, Y_{lm}(\theta_q,\varphi_q) \, {\rm d}\varphi_k \, ,
\end{eqnarray}
where ${\bf q} = {\bf p} - {\bf k}$ and the prefactor $c_{\rm tw}$ is given by
\begin{eqnarray}
  c_{\rm tw} = (-\rmi)^{m_\gamma+l} \, 2^{2l+2} \, n^{-l-2} \, Z^{l+4} \, l! \, \sqrt{\frac{2 \varkappa \, (n-l-1)!}{\pi Z^3 (n+l)!}} \, .
\end{eqnarray}
It follows from these formulas, that the computation of the transition amplitude $M_{fi}^{\rm tw}(\theta_p,\varphi_p)$ is reduced to an integration over the azimuthal angle $\varphi_k$. To perform such an integration \textit{analytically} we need to re--write the integrand in the right--hand side of Eq.~(\ref{eq:OriginalIntegral}) in such a way that its $\varphi_k$--dependence becomes explicit. We start from the product of the polarization ${\bf e}_{{\bf k}, \Lambda}$ and the momentum ${\bf p}$ vectors which can be simplified to
\begin{equation}
  \label{eq:PolarizationVectorExpansion}
  ({\bf e}_{{\bf k}, \Lambda} \cdot {\bf p}) = \frac{p_\perp}{\sqrt{2}} \, c_{-1} \, \rme^{\rmi (\varphi_k - \varphi_p)}
  + p_z \, c_0 - \frac{p_\perp}{\sqrt{2}} \, c_{+1} \, \rme^{-\rmi (\varphi_k-\varphi_p)}
\end{equation}
with
\begin{equation}
  p_\perp = p \sin\theta_p , \; \; p_z = p \cos\theta_p \; .
\end{equation}
As a second step, we shall expand the spherical harmonics $Y_{l, m}(\theta_q, \varphi_q)$ in terms of $Y_{\sigma, \mu}(\theta_k, \varphi_k)$. To perform this expansion, we introduce the solid spherical harmonics:
\begin{equation}
  \mathcal{R}_{l, m}({\bf r}) = \sqrt{\frac{4\pi}{2l+1}} \, r^l \, Y_{l, m}(\theta_r, \varphi_r),
\end{equation}
for which the following addition theorem \cite{tough77} holds
\begin{eqnarray}
\label{eq:AdditionTheorem}
  \mathcal{R}_{l, m}({\bf r} + {\bf a}) = \sum_{\sigma=0}^l \sum_{\mu = - \sigma}^\sigma {2 l \choose 2 \sigma}^{1/2} \nonumber\\
  \times \cgc{\sigma,\mu;l-\sigma,m-\mu}{l m} \mathcal{R}_{\sigma, \mu}({\bf r}) \, \mathcal{R}_{l-\sigma, m-\mu}({\bf a}) \, ,
\end{eqnarray}
where $\cgc{...}{...}$ is a Clebsch--Gordan coefficient and ${\bf a}$ is some displacement vector. By writing
\begin{eqnarray}
q^l \, Y_{l, m}(\theta_q,\varphi_q) = \sqrt{\frac{2l+1}{4\pi}} \, \mathcal{R}_{l, m}({\bf q})  
\end{eqnarray}
with 
\begin{eqnarray}
{\bf q} = {\bf p} - {\bf k} 
\end{eqnarray}
and applying Eq.~(\ref{eq:AdditionTheorem}), we obtain 
\begin{eqnarray}
  q^l \ Y_{lm}(\theta_q,\varphi_q) = \sum_{\sigma=0}^l \sum_{\mu = - \sigma}^\sigma {2 l \choose 2 \sigma}^{1/2} \, \sqrt{\frac{4 \pi (2l+1)}{(2\sigma+1)(2 (l-\sigma)+1)}} \nonumber\\ 
  (-1)^{l-\sigma} p^\sigma k^{l-\sigma} \cgc{\sigma,\mu;l-\sigma,m-\mu}{l m} Y_{\sigma, \, \mu}(\theta_p,\varphi_p) Y_{l-\sigma, \, m-\mu}(\theta_k,\varphi_k) \, .
\end{eqnarray}
By re-writing the spherical harmonics $Y_{l-\sigma, \, m-\mu}(\theta_k,\varphi_k)$ in this expression in terms of $\rme^{\rmi (m-\mu) \varphi_k}$, we get
\begin{eqnarray}
\label{eq:SphericalHarmonicsExpansion}
  q^l \ Y_{lm}(\theta_q,\varphi_q) &=& \sum_{\sigma=0}^l \sum_{\mu = - \sigma}^\sigma h_{\sigma,\mu}(\theta_p,\varphi_p) \, \rme^{\rmi (m-\mu) \varphi_k}
\end{eqnarray}
with
\begin{eqnarray}
  h_{\sigma,\mu}(\theta_p,\varphi_p) &= \sqrt{\frac{2l+1}{2\sigma+1}} \sqrt{\frac{(l-\sigma-m+\mu)!}{(l-\sigma+m-\mu)!}} {2 l \choose 2 \sigma}^{1/2} (-1)^{l-\sigma} p^\sigma k^{l-\sigma} \nonumber \\
  &\times \cgc{\sigma,\mu;l-\sigma,m-\mu}{l m} Y_{\sigma, \mu}(\theta_{p},\varphi_{p}) P_{l-\sigma,m-\mu}(\cos\theta_k) \, .
\end{eqnarray}

\medskip

Having deduced the explicit $\varphi_k$--dependence of the product $({\bf e}_{{\bf k},\Lambda} \cdot {\bf p})$ and of the expression $q^l \ Y_{l, m}(\theta_q,\varphi_q)$ from Eq.~(\ref{eq:OriginalIntegral}), it remains to perform the Fourier expansion
\begin{eqnarray}
  \label{eq:FourierExpansion}
  \frac{1}{(q^2+\delta^2)^{l+2}} \, C_{n-l-1}^{l+1} \left(\frac{q^2-\delta^2}{q^2+\delta^2}\right) = \sum_{\nu=-\infty}^{\infty} f_\nu(\theta_p,\varphi_p) \, \rme^{\rmi \nu \varphi_k},
\end{eqnarray}
of the Gegenbauer polynomial $C_{n-l-1}^{l+1}(x)$ divided by some polynomial. In order to compute the expansion coefficients
\begin{equation}
  \label{ref:FourierCoefficients}
  f_\nu(\theta_p,\varphi_p) = \frac{1}{2\pi} \int_{0}^{2\pi} \frac{1}{(q^2+\delta^2)^{l+2}} \, C_{n-l-1}^{l+1} \left(\frac{q^2-\delta^2}{q^2+\delta^2}\right) \, \rme^{-\rmi \nu \varphi_k} \, {\rm d}\varphi_k \, ,
\end{equation}
we write the Gegenbauer polynomials explicitly \cite{abramowitz70} as
\begin{equation}
  \label{eq:Gegenbauer}
  C_{n-l-1}^{l+1} \left(\frac{q^2-\delta^2}{q^2+\delta^2}\right) = \sum_{\eta=0}^{\left \lfloor(n-l-1)/2 \right \rfloor} \! t_\eta \left (\frac{q^2-\delta^2}{q^2+\delta^2} \right )^{n-l-1-2\eta},
\end{equation}
where $\left \lfloor x \right \rfloor$ is the largest integer not greater than $x$ and
\begin{eqnarray}
  t_\eta = 2^{n-l-1-2\eta} \frac{(-1)^\eta \ (n-\eta-1)!}{\eta! \ l! \ (n-l-1-2\eta)!} \, .
\end{eqnarray}
By inserting Eq.~(\ref{eq:Gegenbauer}) into (\ref{ref:FourierCoefficients}) we find the Fourier expansion coefficients in the form
\begin{eqnarray}
  \label{eq:FourierIntegral}
  f_\nu(\theta_p,\varphi_p) = \frac{1}{2\pi} \sum_{\eta=0}^{\left \lfloor(n-l-1)/2 \right \rfloor} t_\eta \int_{0}^{2\pi} \frac{(q^2-\delta^2)^{n-l-1-2\eta}}{(q^2+\delta^2)^{n+1-2\eta}} \rme^{-\rmi \nu \varphi_k} \, {\rm d}\varphi_k \, .
\end{eqnarray}
In order to compute the integral from above, we perform the substitution $\varphi_k \rightarrow \varphi_p - \varphi_k$ and introduce the new (integration) variable
\begin{eqnarray}
  z = \rme^{\rmi \varphi_k} \, ,
\end{eqnarray}
and, hence, re--write (\ref{eq:FourierIntegral}) as
\begin{eqnarray}
  \label{eq:PathIntegrals}
  f_\nu(\theta_p,\varphi_p) = \frac{- \rmi}{2\pi}\, \alpha^{-l-2} \, \rme^{-\rmi \nu \varphi_p} \, \sum_{\eta=0}^{\left \lfloor(n-l-1)/2 \right \rfloor} t_\eta \int_\chi  g_{\eta,\nu}(z) \, {\rm d}z \, .
\end{eqnarray}
Here, the integration contour $\chi$ is the complex unit circle, and the function $g_{\eta,\nu}(z)$ is given by
\begin{eqnarray}
  \label{eq:g_function}
  g_{\eta,\nu}(z) &=& z^{\nu+l+1} (z-z_1)^{n-l-1-2\eta}(z-z_2)^{n-l-1-2\eta} \nonumber \\
  &\times& (z-z_3)^{-n-1+2\eta} (z-z_4)^{-n-1+2\eta} \, ,
\end{eqnarray}
with
\begin{eqnarray}
  \alpha &=& - p_\perp \varkappa,
\end{eqnarray}
\begin{eqnarray}
  z_1 & = & \left ( 2 p_\perp \varkappa \right )^{-1} \bigg [ p^2+k^2-\delta^2-2 p_z k_z \nonumber \\
  & + & \sqrt{(p^2+k^2-\delta^2-2 p_z k_z)^2-4 p_\perp^2 \varkappa^2 } \, \, \bigg ],
\end{eqnarray}
\begin{eqnarray}
  z_2 & = & \left ( 2 p_\perp \varkappa \right )^{-1} \bigg [ p^2+k^2-\delta^2-2 p_z k_z \nonumber \\
  & - & \sqrt{(p^2+k^2-\delta^2-2 p_z k_z)^2-4 p_\perp^2 \varkappa^2} \, \,  \bigg ],
\end{eqnarray}
\begin{eqnarray}
  z_3 & = & \left ( 2 p_\perp \varkappa \right )^{-1} \bigg [ p^2+k^2+\delta^2-2 p_z k_z \nonumber \\
  & + & \sqrt{(p^2+k^2+\delta^2-2 p_z k_z)^2-4 p_\perp^2 \varkappa^2} \, \,  \bigg ]
\end{eqnarray}
and
\begin{eqnarray}
  \label{eq:Z4}
  z_4 & = & \left ( 2 p_\perp \varkappa \right )^{-1} \bigg [ p^2+k^2+\delta^2-2 p_z k_z \nonumber \\
  & - & \sqrt{(p^2+k^2+\delta^2-2 p_z k_z)^2-4 p_\perp^2 \varkappa^2} \, \,  \bigg ].
\end{eqnarray}
As seen from Eq.~(\ref{eq:g_function}), the $g_{\eta,\nu}(z)$ is a rational function with poles at $z = 0$ and $z = z_4$ in the unit circle of order $-\nu-l-1$ and $n+1-2\eta$, respectively. We can use, therefore, the residue theorem \cite{rudin87} to calculate the integral over $z$ in Eq.~(\ref{eq:PathIntegrals}) analytically
\begin{eqnarray}
  \label{eq:MatrixElementResidueTheorem}
  f_\nu(\theta_p \, , \varphi_p) = \rme^{-\rmi \nu \varphi_p} \alpha^{-l-2} \sum_{\eta=0}^{\left \lfloor(n-l-1)/2 \right \rfloor} t_\eta \,
  \left( {\rm Res}(g_{\eta,\nu},0) + {\rm Res}(g_{\eta,\nu},z_4) \right) \, ,
\end{eqnarray}
where
\begin{eqnarray}
  \label{eq:ResidueTheorem}
  {\rm Res}(g_{\eta,\nu},z_{\rm pole}) = \frac{1}{(k-1)!} \lim_{z \to z_{\rm pole}} \frac{\partial^{k-1}}{\partial z^{k-1}} (z-z_{\rm pole})^k g_{\eta,\nu}(z),
\end{eqnarray}
and $k$ is the order of the pole $z_{\rm pole}$. By inserting the function $g_{\eta,\nu}(z)$ into the right--hand side of Eq.~(\ref{eq:ResidueTheorem}) and evaluating the derivative for the two poles $z=0$ and $z=z_4$, we finally obtain
\begin{eqnarray}
  {\rm Res}(g_{\eta,\nu},0) = \frac{1}{(-\nu-l-2)!} \sum_{i_0=0}^{-\nu-l-2} \ \sum_{i_1=0}^{-\nu-l-2-i_0} \ \sum_{i_2=0}^{-\nu-l-2-i_0-i_1} (-1)^{\nu-l-2} \nonumber \\
  \times {-\nu-l-2 \choose i_0} \, {-\nu-l-2 -i_0 \choose i_1} \, {-\nu-l-2 - i_0 - i_1 \choose i_2} \nonumber \\
  \times (n-l-1-2\eta)_{i_0} \, (n-l-1-2\eta)_{i_1} \nonumber \\
  \times (-n-1+2\eta)_{i_2} \, (-n-1+2\eta)_{-\nu-l-2-i_0-i_1-i_2} \nonumber \\
  \times z_1^{n-l-1-2\eta-i_0} \, z_2^{n-l-1-2\eta-i_1} \, z_3^{-n-1+2\eta-i_2} \nonumber \\
  \times z_4^{-n+1+2\eta+\nu+l+i_0+i_1+i_2} \, ,
\end{eqnarray}
and
\begin{eqnarray}
  \label{eq:Residuez4}
  {\rm Res}(g_{\eta,\nu},z_4) = \frac{1}{(n-2\eta)!} \sum_{i_0=0}^{n-2\eta} \ \sum_{i_1=0}^{n-2\eta-i_0} \ \sum_{i_2=0}^{n-2 \eta-i_0-i_1}  {n-2\eta \choose i_0}  \nonumber \\
  \times {n-2\eta -i_0 \choose i_1} {n-2\eta - i_0 - i_1 \choose i_2} (\nu+l+1)_{i_0} (n-l-1-2\eta)_{i_1}  \nonumber \\
  \times (n-l-1-2\eta)_{i_2} (-n-1+2\eta)_{n-2\eta-i_0-i_1-i_2} \nonumber  \\
  \times z_4^{\nu+l+1-i_0} (z_4-z_1)^{n-l-1-2\eta-i_1} (z_4-z_2)^{n-l-1-2\eta-i_2} \nonumber \\
  \times (z_4-z_3)^{-2n-1+4\eta+i_0+i_1+i_2} \, ,
\end{eqnarray}
where $\left (x \right)_j = x \, (x-1) \, (x-2) \ldots (x-j)$ is the the falling factorial. With the help of Eqs.~(\ref{eq:MatrixElementResidueTheorem})--(\ref{eq:Residuez4}) one may evaluate the expansion coefficients $f_\nu(\theta_p,\varphi_p)$ that enter the Fourier expansion (\ref{eq:FourierExpansion}). Numerical analysis of these coefficients has shown, moreover, that for the ground and low--lying excited atomic states and for small photon energies ($\approx 100 \, \rm{eV}$), the summation over $\nu$ in Eq.~(\ref{eq:FourierExpansion}) may be restricted just to terms with $|\nu| \lesssim 3$.

\medskip

After the separate evaluation of the three parts of the integrand in Eq.~(\ref{eq:OriginalIntegral}), we are ready now to compute the matrix element $M_{fi}^{\rm tw}(\theta_p,\varphi_p)$. That is, by inserting Eqs.~(\ref{eq:PolarizationVectorExpansion}), (\ref{eq:SphericalHarmonicsExpansion}) and (\ref{eq:FourierExpansion}) in the right--hand side of Eq.~(\ref{eq:OriginalIntegral}), we find
\begin{eqnarray}
  \label{eq:RewrittenMatrixElement}
  M_{fi}^{{\rm tw}}(\theta_p,\varphi_p) & = & c_{\rm tw} \sum_{\sigma=0}^l \sum_{\mu = - \sigma}^\sigma \sum_{\nu=-\infty}^{\infty} \int_0^{2 \pi} h_{\sigma,\mu}(\theta_p,\varphi_p) f_\nu(\theta_p,\varphi_p) \nonumber \\
  & \times & \rme^{\rmi (m_\gamma+m+\nu-\mu) \varphi_k} \, \rme^{-\rmi {\bf b} {\bf q} } \nonumber \\ 
  & \times & \left(\frac{p_\perp}{\sqrt{2}} \, c_{-1} \, \rme^{-\rmi (\varphi_p - \varphi_k)} + p_z \, c_0 - \frac{p_\perp}{\sqrt{2}} \, c_{+1} \, \rme^{\rmi (\varphi_p - \varphi_k)}\right) {\rm d} \varphi_k \, .
\end{eqnarray}
We further utilize the well--known integral representation \cite{abramowitz70} of the Bessel functions
\begin{eqnarray}
  \label{eq:BesselFunctionRepresentation}
  \int_0^{2\pi} \rme^{\rmi l \varphi} \rme^{\rmi x \cos(\phi-\varphi)} {\rm d}\varphi = 2 \pi \, \rmi^l \, \rme^{\rmi l \Phi} J_l(x) \, ,
\end{eqnarray}
to analytically perform the integration over the $\varphi_k$-angle in Eq.~(\ref{eq:RewrittenMatrixElement}), and to yield finally the general expression
\begin{eqnarray}
  \label{eq:FinalFormMatrixElement}
  M_{fi}^{\rm tw}(\theta_p,\varphi_p) = 2 \pi \, c_{\rm tw} \, \sum_{\sigma=0}^l \sum_{\mu = - \sigma}^\sigma \sum_{\nu=-\infty}^{\infty} h_{\sigma,\mu}(\theta_p,\varphi_p) f_\nu(\theta_p,\varphi_p) \nonumber \\
  \times \, \rmi^{m_\gamma+m+\nu-\mu} \, \rme^{-\rmi {\bf b}\cdot{\bf p}} \, \rme^{\rmi b_z k_z} \, \rme^{\rmi (m_\gamma+m+\nu-\mu) \varphi_b} \nonumber \\
  \times \, \Big (\rmi \, \frac{p_\perp}{\sqrt{2}} \, c_{-1} \, \rme^{-\rmi (\varphi_p-\varphi_b)} \, J_{m_\gamma+m+\nu-\mu+1}(\varkappa \, b_\perp) \nonumber\\
  \; \; \; \; \; + p_z \, c_0 \, J_{m_\gamma+m+\nu-\mu}(\varkappa \, b_\perp) \nonumber\\ 
  \; \; \; \; \; + \rmi \, \frac{p_\perp}{\sqrt{2}} \, c_{+1} \, \rme^{\rmi (\varphi_p-\varphi_b)} \, J_{m_\gamma+m+\nu-\mu-1}(\varkappa \, b_\perp) \Big) \, ,
\end{eqnarray}
for the amplitude that describes the bound--free electron transition in the field of the twisted (Bessel) light. In this expression, $\varphi_b$ is the azimuthal angle of the impact parameter, which should be taken to be \textit{zero} for the geometry used in the present study (cf.~Fig.~\ref{Fig:GeometryBesselBeam}).

%
%
%
%

\end{document}